\documentclass[preprintnumbers,twocolumn,prl,aps,superscriptaddress,footinbib,amsfonts,amsmath,amssymb,showpacs]{revtex4-1}

\usepackage[T1]{fontenc}
\usepackage{blindtext}

\usepackage{graphics,graphicx}% Include figure files
\usepackage{dcolumn}% Align table columns on decimal point
\usepackage{bm}% bold math
\usepackage{epsfig}
\usepackage[usenames]{color}
\usepackage{hyperref} 
\usepackage{mathbbol}
\usepackage{epstopdf}
\usepackage{simplewick}
\usepackage[utf8]{inputenc} 
\usepackage{slashed}

\def\eq#1{{Eq.~(\ref{#1})}}
\def\fig#1{{Fig.~\ref{#1}}}
\newcommand{\ben}{\begin{eqnarray*}}
\newcommand{\een}{\end{eqnarray*}}

\newcommand{\as}{\alpha_s}

\newcommand{\dhd}{{\textstyle d}
\lower.03ex\hbox{\kern-0.38em$^{\scriptstyle-}$}\kern-0.05em{}}
\newcommand{\dbar}{{\textstyle \delta}
\lower.03ex\hbox{\kern-0.38em$^{\scriptstyle-}$}\kern-0.05em{}}

\newcommand{\env}[1]{\texttt{#1}}
\newcommand{\ul}[1]{\underline{#1}}

\begin{document}

\preprint {LA-UR-16-27995}
\preprint{RBRC-1207}

\title{Small-$x$ asymptotics of the quark helicity distribution}

\author{Yuri V. Kovchegov} 
         \email[Email: ]{kovchegov.1@osu.edu}
         \affiliation{Department of Physics, The Ohio State
           University, Columbus, OH 43210, USA}
\author{Daniel Pitonyak}
	\email[Email: ]{dpitonyak@quark.phy.bnl.gov}
	\affiliation{Division of Science, Penn State University-Berks, Reading, PA 19610, USA}
	\affiliation{RIKEN BNL Research Center, Brookhaven National
          Laboratory, Upton, New York 11973, USA}
\author{Matthew D. Sievert}
  \email[Email: ]{sievertmd@lanl.gov}
	\affiliation{Theoretical Division, Los Alamos National Laboratory, Los Alamos, NM 87545, USA}
	\affiliation{Physics Department, Brookhaven National
          Laboratory, Upton, NY 11973, USA \\}

\begin{abstract}
  We construct a numerical solution of the small-$x$ evolution
  equations recently derived in \cite{Kovchegov:2015pbl} for the
  (anti)quark helicity TMDs and PDFs as well as the $g_1$ structure
  function.  We focus on the case of large $N_c$ where one finds a
  closed set of equations. Employing the extracted intercept, we are
  able to predict directly from theory the behavior of the helicity
  PDFs at small $x$, which should have important phenomenological
  consequences. We also give an estimate of how much of the proton's
  spin may be at small $x$ and what impact this has on the so-called
  ``spin crisis.''
\end{abstract}

\pacs{12.38.-t, 12.38.Bx, 12.38.Cy}

\maketitle

%%%%%%%%%%%%%%%%%%%%%%%%%%%%%%%%%%%%%%%%%%%%%%%%%%%%%%%%%%%%%%%%%%%%%%%%%%%%%%%

%%% Introduction %%%%%%
{\it Introduction} $\phantom{A}$ For many decades, it has been known
that the proton is a complex object composed of quarks, antiquarks,
and gluons (collectively called partons).  The properties of the proton are thus emergent phenomena arising from the dynamics of partons.  For
example, the spin of the proton ($= 1/2$ in units of $\hbar$), which
is one of its most fundamental quantum numbers, should be a sum of the
spin and orbital angular momentum (OAM) of its partons.  This can be
expressed in terms of helicity sum
rules~\cite{Jaffe:1989jz,Ji:1996ek,Ji:2012sj,footnote1}, like
that of Jaffe and Manohar~\cite{Jaffe:1989jz}
\begin{align}
  \label{eq:sum_rule}
  S_q + L_q + S_G + L_G = \frac{1}{2}\,, 
\end{align}
where $S_q$ and $S_G$ are the spin of the quarks and gluons,
respectively, while $L_q$ and $L_G$ denote their OAM.  The
quantities $S_q$ and $S_G$ are defined as the following integrals over Bjorken-$x$ at
a fixed momentum scale $Q^2$,
\begin{align}
  \label{eq:net_spin}
  S_q (Q^2) & = \frac{1}{2} \, \int\limits_0^1 \!dx \, \Delta \Sigma
  (x,
  Q^2)\,, \\
  \label{eq:net_spin2}
   S_G (Q^2) & = \int\limits_0^1 \!dx \, \Delta G (x, Q^2)\,,
\end{align}
with 
\begin{align}
  \label{eq:Sigma}
  \Delta \Sigma (x, Q^2) = \left[ \Delta u + \Delta {\bar u} + \Delta
    d + \Delta {\bar d} + \ldots \right] \! (x, Q^2)\,,
\end{align}
where the helicity parton distribution functions (PDFs) for a parton
of flavor $f = u,\,\bar{u},\,d,\,\bar{d},\dots,\,G$ are denoted by
$\Delta f$.

In the late 1980s, the community was largely surprised when the
European Muon Collaboration (EMC) measured $S_q$ to be a significantly
smaller fraction of the proton's spin than had been na\"{i}vely
expected~\cite{Ashman:1987hv,Ashman:1989ig}.  This result triggered
the so-called ``spin crisis'' centered around the question of how the pieces in
Eq.~(\ref{eq:sum_rule}) add up to 1/2.  To help pin down
another term in this sum, there has been intense effort over the last
decade to measure and extract $S_G$.  Recent experiments show that
$S_G$ can give a more substantial fraction of the proton's spin than
once thought~\cite{Adamczyk:2014ozi,Adare:2015ozj}.  The current quark
and gluon spin values extracted from the experimental data are $S_q
(Q^2 = 10\, \mbox{GeV}^2) \approx 0.15 \div 0.20$ (integrated over
$0.001 < x <1$) and $S_G (Q^2 = 10 \, \mbox{GeV}^2) \approx 0.13 \div
0.26$ (integrated over $0.05 < x <1$)~\cite{footnote2}.  Conventional
wisdom, then, is that the rest of the proton's spin is due to quark
and gluon OAM.  However, note that the quoted values for $S_q$ and
$S_G$ are for integrals over a truncated range $x_{min} < x< 1$ (where
the relevant quantities are constrained by data), while the formulae
in Eqs.~(\ref{eq:net_spin}), (\ref{eq:net_spin2}) involve integrals over the full range $0 <
x < 1$.  This leaves open the possibility that there could be significant quark
and gluon spin at small $x$, which is the scenario we will explore in this
Letter.

The use of the small-$x$ formalism to analyze quark polarization was
pioneered decades ago by Kirschner and Lipatov~\cite{Kirschner:1983di}
(see also~\cite{Kirschner:1994rq,Kirschner:1994vc,Griffiths:1999dj})
and later by Bartels, Ermolaev, and Ryskin (herein referred to as BER)
in the context of the structure function
$g_1(x,Q^2)$~\cite{Bartels:1995iu,Bartels:1996wc}.  In particular, BER
resummed double logarithms $\as \ln^2 (1/x)$ using infrared evolution
equations to predict a strong {\it growth} in $g_1(x,Q^2)$ at
small-$x$, a scenario that would have a major impact on the spin
crisis.  In a recent work, we formulated the problem in a different
language, which employs light-cone Wilson line operators and color
dipoles~\cite{Mueller:1994rr,Mueller:1994jq,Mueller:1995gb,Balitsky:1996ub,Balitsky:1998ya,Kovchegov:1999yj,Kovchegov:1999ua,Jalilian-Marian:1997dw,Jalilian-Marian:1997gr,Iancu:2001ad,Iancu:2000hn},
to derive evolution equations relevant for the (collinear and
transverse momentum dependent (TMD)) helicity PDFs as well as the
$g_1$ structure function~\cite{Kovchegov:2015pbl}.

In what follows, we solve these helicity evolution equations
numerically (in the limit of a large number of colors $N_c$) in
order to give a direct input from theory on the small-$x$ behavior of
helicity PDFs, which should have important phenomenological
consequences. We extract the high-energy intercept $\alpha_h$ to
predict the small-$x$ asymptotics of $\Delta\Sigma(x,Q^2) \sim
(1/x)^{\alpha_h}$ and estimate how much of the proton's spin one can
expect to find at low $x$.

%%% The helicity evolution equations %%%
{\it The helicity evolution equations} $\phantom{A}$ 
As shown in \cite{Kovchegov:2016zex}, at small $x$ the quark helicity PDF in the
flavor-singlet case
$\Delta q^S(x,Q^2)$ (and, therefore, $\Delta\Sigma(x,Q^2)$) can be written in terms of 
the impact-parameter integrated polarized dipole amplitude
$G(x_{10}^2,z)$
as~\cite{footnote3}
\begin{align}
  \label{eq:Deltaq}
  \Delta q^S (x, Q^2) = \frac{N_c} { 2 \pi^3} \sum_f\int\limits_{z_i}^1
  \frac{dz}{z} \int\limits_{\frac{1}{z \, s}}^{\frac{1}{zQ^2}}
  \frac{dx_{10}^2}{x_{10}^2} \,G(x_{10}^2,z)\,.
\end{align}
Here
$\ul{x}_{10} = \ul{x}_1 - \ul{x}_0$ is the dipole size, 
% $\ul{b} = (\ul{x}_0 + \ul{x}_1)/2$ is the impact parameter, 
$z$ is the fraction of the probe's longitudinal momentum carried by the
softest (anti)quark in the dipole, $z_i = \Lambda^2/s$, with $\Lambda$
an infrared (IR) momentum cutoff, and $s$ is the center-of-mass energy
squared.  The singularity at $\ul{x}_{10} = 0$ is regulated by
requiring that $x_{10} \equiv |\ul{x}_{10}| > 1/(z \, s)$, with $1/(z
\, s)$ the shortest distance (squared) allowed in the problem.

To determine $G(x_{10}^2,z)$, we will solve the evolution equations
derived in \cite{Kovchegov:2015pbl}. They resum powers of $\as
\ln^2(1/x)$, which is the double-logarithmic approximation
(DLA). Similar to the unpolarized
case~\cite{Mueller:1994rr,Mueller:1994jq,Mueller:1995gb,Balitsky:1996ub,Balitsky:1998ya,Kovchegov:1999yj,Kovchegov:1999ua,Jalilian-Marian:1997dw,Jalilian-Marian:1997gr,Iancu:2001ad,Iancu:2000hn},
helicity evolution equations do not close in general, forming a closed
set only in the large-$N_c$ and large-$N_c \, \& N_f$ limits (with $N_f$
the number of flavors)~\cite{Kovchegov:2015pbl}.
Ignoring leading-logarithmic (LLA) saturation
corrections~\cite{footnote4}, the large-$N_c$ DLA evolution of
$G(x_{10}^2,z)$ is governed by Eq.~(83a) in \cite{Kovchegov:2015pbl}
integrated over all impact parameters,
\begin{align}
  G(x_{10}^2,z) \!=\! \,\,& G^{(0)} (x_{10}^2,z)+ \frac{\as \, N_c}{2
    \pi} \int\limits_{\tfrac{1}{x_{10}^2 s}}^z \frac{d z'}{z'} \, \int\limits_{ \frac{1}{z'
    s}}^{x_{10}^2}
  \frac{d x_{21}^2}{x_{21}^2}  \nonumber  \\
  & \times \, \left[ \Gamma(x_{10}^2,x_{21}^2,z') + 3 \,
    G(x_{21}^2,z') \right]. \label{e:G}
\end{align}
In \eq{e:G}, one also has the object $\Gamma(x_{10}^2,x_{21}^2,z')$,
called a ``neighbor'' dipole amplitude ~\cite{Kovchegov:2015pbl}.
The neighbor dipole obeys the (large-$N_c$, strictly DLA) evolution
equation~\cite{Kovchegov:2015pbl}
\begin{align}
  & \Gamma(x_{10}^2,x_{21}^2,z') = \Gamma^{(0)}(x_{10}^2,x_{21}^2,z')
  + \frac{\as \, N_c}{2 \pi} \int\limits_{\tfrac{1}{x_{10}^2 s}}^{z'} 
	\frac{d z''}{z''} \label{e:Gam} \\
  & \times \! \! \! \int\limits_{\frac{1}{z''  s}}^{\mbox{min} \left\{
      x_{10}^2 , x_{21}^2 \frac{z'} {z''} \right\}} \frac{d
    x_{32}^2}{x_{32}^2} \, \left[ \Gamma(x_{10}^2,x_{32}^2,z'') + 3 \,
    G (x_{32}^2,z'') \right]. \nonumber
\end{align}
Note that in Eqs.~\eqref{e:G}, \eqref{e:Gam} we have neglected
small differences in the dipole sizes $x_{10}^2 \approx x_{20}^2
\approx x_{30}^2$.  The solution to the simultaneous equations
(\ref{e:G}), (\ref{e:Gam}), which we discuss in the next section,
allows us to determine the small-$x$ behavior of $G(x_{10}^2,z)$, and,
hence, of $\Delta q^S (x, Q^2)$ in the dominant flavor singlet channel.

%%% Numerical solution to the large-N_c evolution equations %%%
{\it Numerical solution to the large-$N_c$ evolution equations}
$\phantom{A}$ We start by defining new coordinates,
\begin{align}
  &\hspace{1.3cm}\eta \equiv \ln \frac{z}{z_i} \, , \ \ \  \eta'\equiv \ln \frac{z'}{z_i} \,, \ \ \  \eta''\equiv \ln \frac{z''}{z_i} \, ,  \\[0.3cm]
  & s_{10} \equiv \ln \frac{1}{x_{10}^2 \Lambda^2}\,, \ \ \ s_{21}
  \equiv \ln \frac{1}{x_{21}^2 \Lambda^{2}}\,, \ \ \ s_{32} \equiv \ln
  \frac{1}{x_{32}^2 \Lambda^{2}}, \nonumber
\end{align}
as well as rescaling all $\eta$'s and $s_{ij}$'s, 
\begin{align}\label{resc}
  \eta \to \sqrt{\frac{2\pi}{\as N_c}} \ \eta\,, \ \ \ \ \ s_{ij} \to
  \sqrt{\frac{2\pi}{\as N_c}} \ s_{ij}\,.
\end{align}
Using these variables, we write the large-$N_c$ helicity evolution
equations (\ref{e:G}), (\ref{e:Gam}) as 
\begin{subequations}\label{GGeqns}
\begin{align} 
  & G (s_{10}, \eta) = G^{(0)} (s_{10}, \eta) + \int\limits_{s_{10}}^\eta d \eta' \int\limits_{s_{10}}^{\eta'} d s_{21} \label{e:G_redef} \\
  & \hspace{1.77cm} \times\,\left[ \Gamma (s_{10}, s_{21}, \eta') + 3 \, G (s_{21}, \eta') \right] \notag  \\ 
  & \Gamma (s_{10}, s_{21}, \eta') = \Gamma^{(0)} (s_{10}, s_{21}, \eta') + \int\limits_{s_{10}}^{\eta'} d \eta'' \label{e:Gam_redef} \\
  & \times\int\limits_{\mbox{max} \left\{ s_{10}, s_{21} + \eta'' -
      \eta' \right\}}^{\eta''} \hspace*{-1cm} d s_{32} \left[ \Gamma
    (s_{10}, s_{32}, \eta'') + 3 \, G (s_{32}, \eta'') \right]. \notag
\end{align}
\end{subequations}
Note that the ranges of the $s_{21}$ and $s_{32}$ integrations are
restricted to positive values of $s_{21}$ and $s_{32}$ as long as
$s_{10}$ is positive; therefore, we always stay above the IR cutoff
$\Lambda$ (in momentum space).  The initial conditions for
Eqs.~\eqref{GGeqns} are %(\ref{e:G_redef}), (\ref{e:Gam_redef}) are
\cite{Kovchegov:2016zex,footnote5}
\begin{align}
  G^{(0)}\! (s_{10}, \eta) &=  \Gamma^{(0)} \!(s_{10}, s_{21}, \eta) \nonumber\\
  &=  \as^2\pi \tfrac{C_F}{N_c} \!\left[ C_F \, \eta -2 (\eta -
    s_{10}) \right] ,\label{e:in_cond}
\end{align}
with $C_F = (N_c^2-1)/(2 N_c)$.  Since the equations at hand are linear, and we are
mainly interested in the high-energy intercept, we can scale out
$\as^2\pi\,C_F/N_c$.

In order to solve Eqs.~\eqref{GGeqns}~\cite{footnote6}, we first write down a
discretized version of them
\begin{subequations}\label{eqs_discr}
\begin{align}
  G_{ij} & \!=\! G^{(0)}_{ij} + \Delta \eta\,\Delta s \sum_{j'=i}^{j-1} \sum_{i' = i}^{j'}  \left[ \Gamma_{ii'j'} + 3 \, G_{i'j'} \right], \label{e:G_dis}\\
  \Gamma_{ikj} & \!=\! \Gamma^{(0)}_{ikj} + \Delta \eta\,\Delta s
  \!\sum_{j'=i}^{j-1} \ \sum_{i' = \mbox{max} \{ i, k+j'-j \}
  }^{j'}\!\! \left[ \Gamma_{ii'j'} + 3 \,
    G_{i'j'}\right],\nonumber\\ \label{e:Gam_dis}
\end{align}
\end{subequations}
where $G_{ij} \equiv G(s_i,\eta_j)$, $\Gamma_{ikj} \equiv \Gamma(s_i,
s_k, \eta_j)$, and
\begin{align}
\Delta \eta = \frac{\eta_{max}}{N_\eta}\,, \ \ \  \Delta s = \frac{\, s_{max}}{\, N_s} \,,
\end{align}
with $\eta_{max}$ the maximum $\eta$ value and $N_\eta$ the number of
grid steps in the $\eta$ direction, and likewise for $s_{max}$, $N_s$.
The discretized equations \eqref{eqs_discr} 
are exact in the limit $\Delta\eta\,,\,\Delta s\to 0$ and
$\eta_{max}\,,\,s_{max} \!\to \infty$.  To optimize the numerics, we set 
$\eta_{max} = s_{max}$.

\begin{figure}[tb]
\centering \includegraphics[width=0.47 \textwidth]{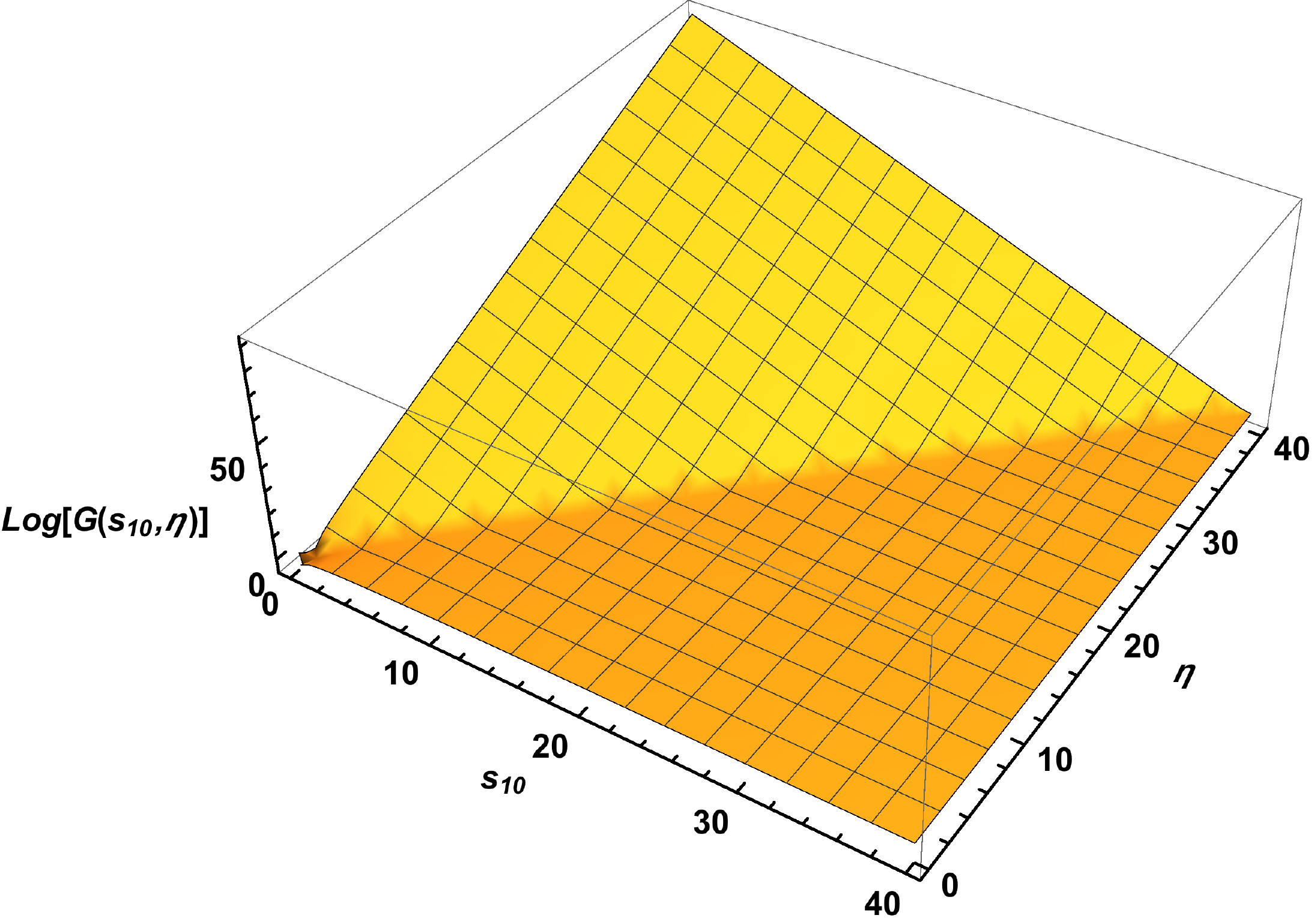}
\caption{The numerical solution of Eqs.~\eqref{GGeqns} for the
  polarized dipole amplitude $G$ plotted as a function of rescaled
  ``rapidity'' $\eta$ and transverse variable $s_{10}$.}
\label{f:G}
\end{figure}

With the discretized evolution equations \eqref{eqs_discr} in hand
(along with the initial conditions (\ref{e:in_cond}) suitably
discretized), we first choose values for $\eta_{max} = s_{max}$ and
$\Delta \eta = \Delta s$.  We then systematically go through the
$\eta$-$s$ grid in such a way that each $G_{ij}$ (and $\Gamma_{ikj}$)
only depends on $G,\Gamma$ values that have already been calculated.
Thus, we can determine $G_{ij}$ for each $i,j$. Our numerical solution
(for $\eta_{max} = 40$, $\Delta \eta = 0.05$) is plotted in \fig{f:G}.
%%%%%%%%%%%%%%%
%%%%%%%%%%%%%%%

\begin{figure}[h]
\centering \includegraphics[scale=0.45]{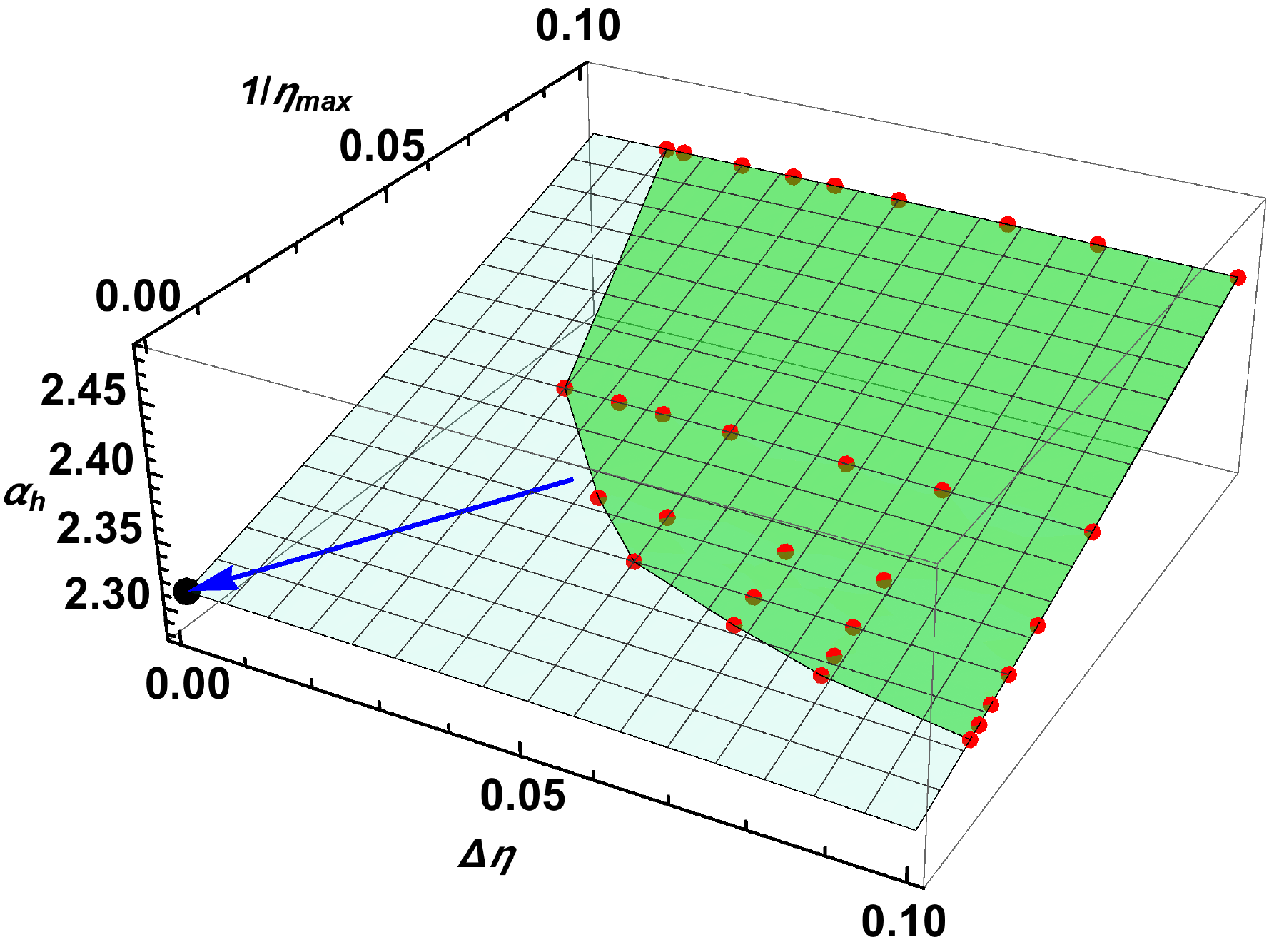} 
\caption{Numerical results for our extraction of $\alpha_h$.  The
  ``data'' points are the intercepts we obtained for various
  $(\eta_{max},\Delta\eta)$.  The plane gives our model with the best
  fit to the data, namely, $\alpha_h(\eta_{max},\Delta\eta) =
  -0.54(\Delta\eta)^2+0.063(\Delta\eta)(1/\eta_{max})+0.0027(\Delta\eta)+1.53(1/\eta_{max})^2+1.12(1/\eta_{max})+2.31$.
  The dark shaded piece indicates the region that is within our
  computational range, while the light area shows our extrapolation to
  the physical point $1/\eta_{max}=\Delta\eta=0$ (large solid dot).}
\label{f:intercepts}
\end{figure}

We next assume that in the high-energy limit
\begin{align} 
  \label{e:heplimit}
  \hspace*{-1mm} G(s_{10},& \, \eta;\eta_{max},\Delta\eta) \sim
  e^{\alpha_h(\eta_{max},\Delta\eta)\,\eta +
    \beta_h(\eta_{max},\Delta\eta)\, s_{10}}
\end{align}
with some coefficients $\alpha_h$, $\beta_h$ that are functions of
$(\eta_{max},\Delta\eta)$.  We then fit
$\ln\!\left[G(s_{01},\eta;\eta_{max},\Delta\eta)\right]$ vs.~$\eta$
for $s_{10} = 0$, using only $\eta \in [0.75 \, \eta_{max},
\eta_{max}]$.
This allows us to extract the intercept
$\alpha_h(\eta_{max},\Delta\eta)$.  We perform this procedure for
$\eta_{max} = 10,\,20,\,30,\,40,\,50,\,60,\,70$ and $\Delta\eta \in
[\Delta\eta_{min},\,0.1]$, where $\Delta\eta_{min}$ is the smallest
value of $\Delta\eta$, for a given $\eta_{max}$, that is within our
computational limits.  The various intercepts we obtained are shown by
the ``data'' points in Fig.~\ref{f:intercepts}.
%%%%%%%%%%%%%%%
%%%%%%%%%%%%%%%

\begin{figure}[b]
\centering \includegraphics[scale=0.313]{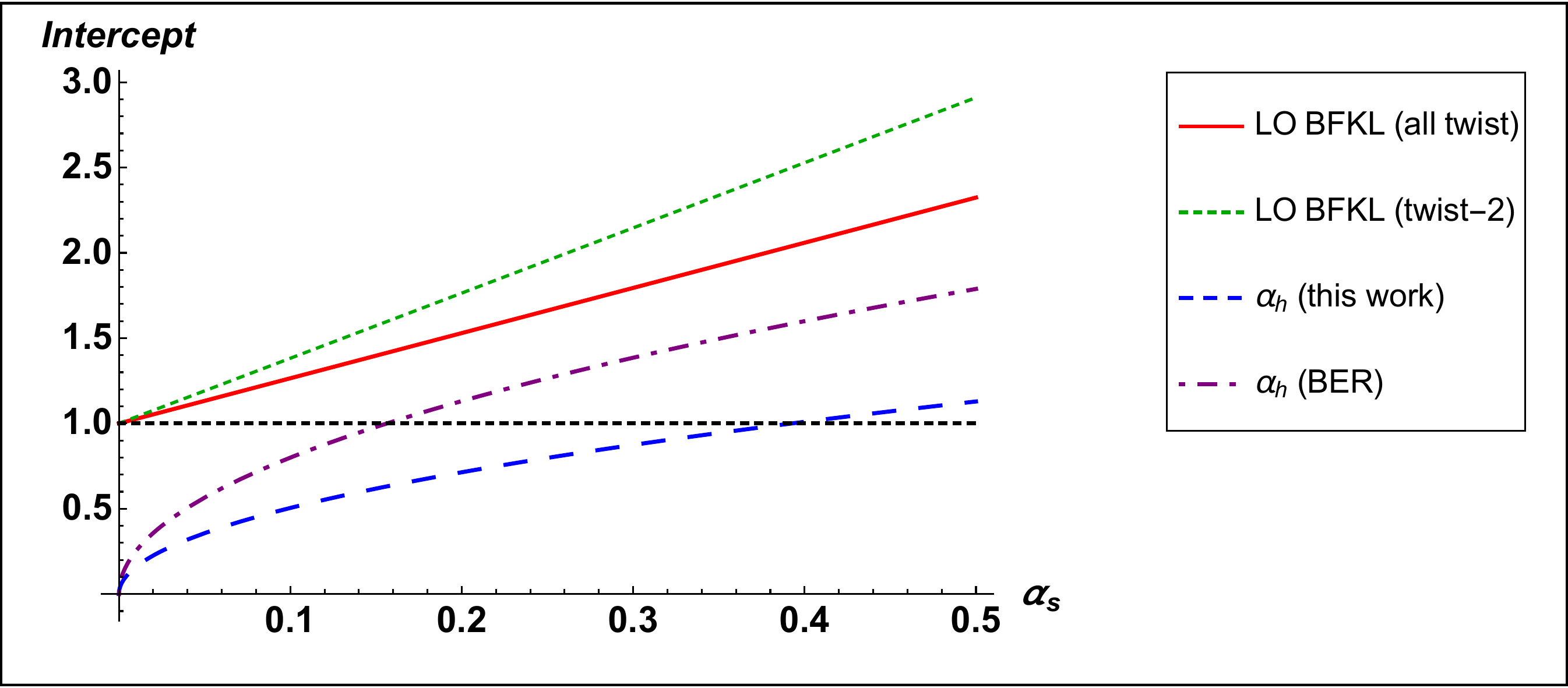} 
%\centering \includegraphics[scale=0.313]{intercepts_comp_old.pdf} 
\caption{Plot of the intercept vs.~$\alpha_s$ for helicity evolution
  (long-dashed and dot-dashed lines) and unpolarized LO BFKL evolution
  (solid and short-dashed lines).  The long-dashed line shows the value of $\alpha_h$
  extracted in this work for large $N_c$ while the dot-dashed line
  gives that for the ``pure glue'' case of BER~\cite{Bartels:1996wc}.} \label{f:inter_comp}
\end{figure}

As a last step, we must extrapolate to the physical point
$\eta_{max}\!\to \infty$, $\Delta\eta\to 0$.  To do this, we first
perform a two-dimensional fit to $\alpha_h(\eta_{max},\Delta\eta)$
using 8 different functional forms (polynomials and powers of $\Delta \eta$ 
and $(1/\eta_{max})$), from which we
can extract $\alpha_h(\eta_{max}\to\infty,\Delta\eta\to 0)\equiv
\alpha_h$ (see Fig.~\ref{f:intercepts}).  Next, we calculate the
corrected Akaike information criterion (AICc) value~\cite{Akaike:1974}
for each curve, which allows us to compare the models against each
other.  (We note that all the fits have $R^2$ values equal to (or
extremely close to) 1.)  Finally, using these AICc values, we compute
a weighted average of $\alpha_h$ from all the fits.  In the end, we
obtain $\alpha_h = 2.31$.  Therefore, we find
\begin{align} 
  \label{e:DeltaSigma}
  \Delta q^S(x,Q^2) \sim \Delta\Sigma(x, Q^2) \sim \left(\frac{1}
    {x}\right)^{\!\alpha_{h}}
\end{align}
with
\begin{align} 
  \label{e:alphah}
  \alpha_{h} = 2.31 \, \sqrt{\frac{\as N_c} {2\pi}}\, ,
\end{align}
where we have reinstated the factor 
$\sqrt{\as N_c/2\pi}$ originally
scaled out by Eq.~(\ref{resc}).  (We also note that $\beta_h \approx - \alpha_h$.) We mention that the uncertainty in
$\alpha_h$ due to the choice of initial conditions and the
extrapolation to the physical point are both $< 1\%$ and negligible.  
We note that the value in Eq.~(\ref{e:alphah}) is
in disagreement with the ``pure glue'' intercept of $3.66\sqrt{\as
  N_c/2\pi}$~\cite{footnote7} obtained by BER \cite{Bartels:1996wc} by about $35\%$.  
In Fig.~\ref{f:inter_comp} we compare
these two intercepts along with that for unpolarized LO BFKL
evolution (all twist and twist-2).  
%%%%%%%%%%%%%%%
%%%%%%%%%%%%%%%
Interestingly, the leading twist approximation to $\alpha_P - 1$ in
BFKL evolution is larger than the exact all-twist intercept by about
$30\%$ \cite{Kovchegov:2012mbw}; it is possible something similar
is occurring for helicity evolution.  In Ref.~\cite{Kovchegov:2016zex}, we have explored this possibility, performed various analytical cross-checks of our helicity evolution equations, and compared to BER where
possible; we have not found any inconsistencies in our result.

%%% Impact on the proton spin %%%
{\it Impact on the proton spin} $\phantom{A}$ In order to determine 
the quark and gluon spin based on Eqs.~(\ref{eq:net_spin}),
(\ref{eq:net_spin2}), one needs to extract the helicity PDFs.  There
are several groups who have performed such analyses, e.g.,
DSSV~\cite{deFlorian:2009vb,deFlorian:2014yva},
JAM~\cite{Jimenez-Delgado:2013boa,Sato:2016tuz},
LSS~\cite{Leader:2005ci,Leader:2010rb,Leader:2014uua},
NNPDF~\cite{Ball:2013lla,Nocera:2014gqa}.  While the focus at small
$x$ has been on the behavior of $\Delta G(x,Q^2)$, there is actually
quite a bit of uncertainty in the size of $\Delta\Sigma(x,Q^2)$ in
that regime as well.

Let us define the truncated integral
\begin{align} 
  \label{e:run_int}
  \Delta\Sigma^{[x_{min}]}\!(Q^2)\equiv \int_{x_{min}}^1 \!dx
  \,\Delta\Sigma(x,Q^2)\,.
\end{align}
One finds for DSSV14~\cite{deFlorian:2014yva} that the central value
of the full integral $\Delta\Sigma^{[0]}\!(10\,{\rm GeV^2})$ is about
$40\%$ smaller than $\Delta\Sigma^{[0.001]}\!(10\,{\rm GeV^2})$.  The
NNPDF14~\cite{Nocera:2014gqa} helicity PDFs lead to a similar
decrease, although, due to the nature of neural network fits, the
uncertainty in this extrapolation is $100\%$.  On the other hand, for
JAM16~\cite{Sato:2016tuz} helicity PDFs the decrease from the
truncated to the full integral of $\Delta\Sigma(x,Q^2)$ seems to be at
most a few percent.  The origin of this uncertainty, and more
generally the behavior of $\Delta\Sigma(x,Q^2)$ at small $x$, is
mainly due to varying predictions for the size and shape of the sea
helicity PDFs, in particular $\Delta
s(x,Q^2)$~\cite{deFlorian:2009vb,deFlorian:2014yva,Jimenez-Delgado:2013boa,Sato:2016tuz,Ball:2013lla,Nocera:2014gqa,Aschenauer:2015ata}.
So far, the only constraint on $\Delta s(x,Q^2)$, and how it evolves
at small $x$, comes from the weak neutron and hyperon decay constants.
Therefore, there is a definite need for direct input from theory on
the small-$x$ intercept of $\Delta\Sigma(x,Q^2)$: this is what we have
provided in this Letter.

\begin{figure}[t]
\centering \includegraphics[scale=0.3]{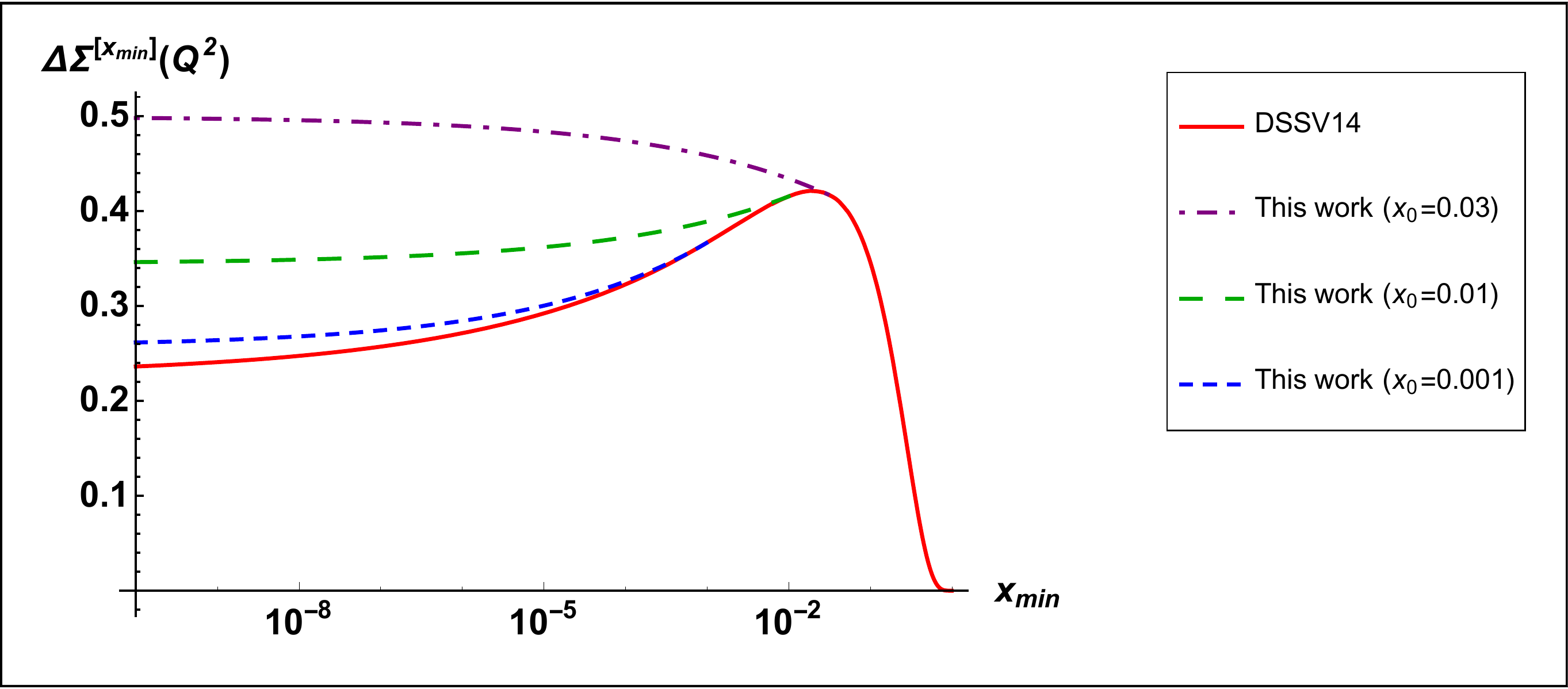} 
\caption{Plot of $\Delta\Sigma^{[x_{min}]}\!(Q^2)$ vs.~$x_{min}$ at
  $Q^2=10\,{\rm GeV}^2$.  The solid curve is from
  DSSV14~\cite{deFlorian:2014yva}.  The dot-dashed, long-dashed, and
  short-dashed curves are from various small-$x$ modifications of
  $\Delta\Sigma(x,Q^2)$ at $x_0=0.03,\,0.01,\,0.001$, respectively,
  using our helicity intercept (see the text for
  details).} \label{f:run_spin}
\end{figure}

We now will attempt to quantify how the small-$x$ behavior of
$\Delta\Sigma(x,Q^2)$ derived here affects the integral in
Eq.~(\ref{eq:net_spin}).  We take a simple approach and leave a more
rigorous phenomenological study for future work.  First, we attach a
curve $\Delta\tilde{\Sigma}(x,Q^2) = N\,x^{-\alpha_h}$ (with
$\alpha_h$ given in (\ref{e:alphah})) to the DSSV14 result for
$\Delta\Sigma(x,Q^2)$ at a particular small-$x$ point $x_0$.  Next, we
fix the normalization $N$ by requiring $\Delta\tilde{\Sigma}(x_0,Q^2)
= \Delta\Sigma(x_0,Q^2)$.  Finally, we calculate the truncated
integral \eqref{e:run_int} of the modified quark helicity PDF
\begin{align} 
  \label{e:run_int_mod}
  \Delta\Sigma_{mod}(x,Q^2) \equiv & \ \theta(x-x_{0}) \,
  \Delta\Sigma(x,Q^2) \notag \\ & + \theta(x_0-x) \,
  \Delta\tilde{\Sigma}(x,Q^2)
\end{align}
for different $x_0$ values. The results are shown in
Fig.~\ref{f:run_spin} for $Q^2 = 10\,{\rm GeV^2}$ and $\alpha_s\approx
0.25$, in which case $\alpha_h \approx 0.80$.

%%%%%%%%%%%%%%%
%%%%%%%%%%%%%%%
We see that the small-$x$ evolution of $\Delta\Sigma(x,Q^2)$ could
offer a moderate to significant enhancement to the quark spin,
depending on where in $x$ the effects set in and on the parameterization 
of the helicity PDFs at higher $x$.  Thus, it will be
important to incorporate the results of this work, and more generally
the small-$x$ helicity evolution equations discussed here, into future
extractions of helicity PDFs.

%%% Conclusion %%%
{\it Conclusion} $\phantom{A}$ In this Letter we have numerically
solved the small-$x$ helicity evolution equations of
Ref.~\cite{Kovchegov:2015pbl} in the large-$N_c$ limit.  We found an
intercept of $\alpha_h=2.31\sqrt{\as N_c/2\pi}$, which, from
Eq.~(\ref{e:DeltaSigma}), is a direct input from theory on the
behavior of $\Delta\Sigma(x,Q^2)$ at small $x$.  Although a more
rigorous phenomenological study is needed, we demonstrated in a simple
approach that such an intercept could offer a moderate to significant
enhancement of the quark contribution to the proton spin.  Therefore,
it appears imperative to include the effects of the small-$x$ helicity
evolution discussed here in future fits of helicity PDFs, especially
those to be obtained at an Electron-Ion Collider.

\acknowledgments We thank S.~Mukherjee for useful discussions on
analyzing the numerical results. We thank R. Sassot and W. Vogelsang
for providing us with the parameters and Fortran code of the DSSV14
fit.  We are also grateful to R.~Furnstahl, A.~Metz, B.~Schenke, and J. Stapleton for useful
conversations.  This material is based upon work supported by the
U.S. Department of Energy, Office of Science, Office of Nuclear
Physics under Award Number DE-SC0004286 (YK) and within the 
framework of the TMD Topical Collaboration (DP) and DOE Contract 
No.~DE-SC0012704 (MS).  DP also received support from the 
RIKEN BNL Research Center.
MS received additional support from an EIC program 
development fund from BNL and from the U.S. Department of Energy, Office of Science under the DOE Early Career Program. \\

%\bibliography{references}

\begin{thebibliography}{99}
\bibitem{footnote5} Note that
  $G(x_{10}^2,z)$ 
  has a sign opposite to that of the spin asymmetry defined in
  Eq.~(A9) of Ref.~\cite{Kovchegov:2015pbl}.  This is analogous to the
  unpolarized case, where the Wilson line correlator is equal to $1 -
  N(x_{10}^2,z)$, where $N(x_{10}^2,z)$ is the imaginary part of the forward
  dipole--target scattering amplitude and has the same sign as the
  total cross section.
\bibitem{footnote7} This intercept is for finite $N_c$, but we also
  checked that the value does not change in the limit of large $N_c$.
\end{thebibliography}

\end{document}